\documentclass[conference]{IEEEtran}
\IEEEoverridecommandlockouts

\usepackage{times}
\usepackage{amsmath,amssymb,amsfonts}
\usepackage{graphicx}
\usepackage{cite}
\usepackage{booktabs}
\usepackage{xcolor}
\usepackage{hyperref}
\def\BibTeX{{\rm B\kern-.05em{\sc i\kern-.025em b}\kern-.08em
    T\kern-.1667em\lower.7ex\hbox{E}\kern-.125emX}}

\begin{document}

\title{On the Practical Performance of Noise Modulation for Ultra-Low-Power IoT: Limitations, Capacity, and Energy Trade-offs\\
\thanks{This work has been partially funded by the xGMobile Project (XGM-AFCCT-2025-8-1-1, XGM-AFCCT-2024-9-1-1) with resources from EMBRAPII/MCTI (Grant 052/2023 PPI IoT/Manufatura 4.0), by CNPq (302085/2025-4, 306199/2025-4), FAPEMIG (PPE-00124-23, RED-00194-23, APQ-04523-23, APQ-05305-23, APQ-03162-24), and by FINEP (nº 1060/2 contract 01.25.0883.00).}
\vspace{-10pt}
}

\author{Felipe A. P. de Figueiredo, Pedro M. R. Pereira, Hadi Zayyani, Evandro C. Vilas Boas, Fernando D. A. García, \\ Hugerles S. Silva, and Rausley A. A. de Souza\thanks{The authors are with the National Institute of Telecommunications (Inatel), Santa Rita do Sapucaí, MG  37536-00, Brazil ([felipe.figueiredo, pedro.marcio, evandro.cesar, fernando.almeida, rausley]@inatel.br, [hadi.zayyani,hugerles.sales]@posdoc.inatel.br)}
\vspace{-20pt}
}

\maketitle

\begin{abstract}
Noise modulation (NoiseMod) has emerged as a promising communication technique for ultra-low-power Internet-of-Things (IoT) devices due to its oscillator-free architecture and non-coherent energy detection. This paper presents a comprehensive analytical and simulation-based evaluation of complex-baseband NoiseMod under AWGN and Rayleigh fading channels. Closed-form expressions for the optimal detection threshold and BER in AWGN are derived, and the Rayleigh fading performance is analyzed together with a practical channel-state-information-free two-antenna selection diversity scheme. A hardware-aware energy model and a Binary Symmetric Channel (BSC)-based capacity analysis are also developed. Results show that although NoiseMod achieves superior BER when compared on a per-sample SNR basis, a fair comparison in terms of $E_b/N_0$ reveals an energy penalty of approximately 11.7 dB relative to coherent BPSK at a BER of $10^{-3}$. Moreover, the required oversampling limits the spectral efficiency to $1/N$ bits per sample. The proposed CSI-free diversity scheme substantially mitigates the performance degradation under Rayleigh fading, while the energy analysis identifies crossover distances beyond which coherent BPSK becomes more energy-efficient. These results clarify the practical trade-offs between hardware simplicity, communication reliability, energy efficiency, and spectral efficiency, providing design guidelines for ultra-low-power IoT systems. The full simulation code is publicly available\footnote{\url{https://github.com/zz4fap/NoiseMod-for-ULP-IoT}}.
\end{abstract}


\vspace{-5pt}
\section{Introduction}
The proliferation of the IoT envisions billions of interconnected devices operating seamlessly in environments where energy harvesting or decades-long battery life is mandatory \cite{b1}. Traditional radio frequency (RF) architectures, employing power-hungry local oscillators, high-resolution analog-to-digital converters (ADCs), and complex modulation schemes, fundamentally bottleneck the scaling of these networks \cite{b2}. Consequently, the research community has pivoted towards ultra-low-power (ULP) communication schemes to relax hardware constraints.

\textit{Noise Modulation} (NoiseMod), also known as variance-shift keying, conveys information through the variance of a transmitted Gaussian noise signal rather than a coherent carrier \cite{b5,b6}. Detection is entirely non-coherent, relying only on received-energy estimation and requiring neither carrier synchronization nor channel estimation \cite{b7}.

Although NoiseMod has been theoretically investigated \cite{b6,b8,b9}, its practical performance relative to conventional schemes such as Non-Coherent Frequency Shift Keying (NC-FSK) or coherent Binary Phase Shift Keying (BPSK) remains insufficiently understood. In particular, comparisons in terms of Bit Error Rate (BER), energy efficiency, channel capacity, and Rayleigh-fading robustness are still lacking.

This paper presents a comprehensive analysis of Noise Modulation. The main contributions are:
\begin{enumerate}
    \item A closed-form Likelihood Ratio Test (LRT) threshold and BER expression for AWGN and Rayleigh channels. This work extends the real-valued energy-detection analysis of \cite{b6} to the complex baseband case.
    \item Demonstration that a 2-antenna selection-diversity scheme can be implemented \emph{without} instantaneous channel-state information (CSI) at essentially no performance cost relative to a genie-aided bound.
    \item A detailed, fully parameterized energy consumption and link-budget model separating static baseband and ADC power, revealing a BER, $E_b/N_0$, energy trade-off and an explicit energy crossover distance against BPSK.
    \item A Binary-Symmetric-Channel (BSC)-based capacity comparison showing the throughput bottleneck caused by oversampling compared to coherent BPSK and non-coherent FSK.
\end{enumerate}

\vspace{-5pt}
\section{System Model and Theoretical Derivations}

\subsection{Noise Modulation Principle in Complex AWGN}
Consider a binary communication system where the transmitter communicates a message bit $b \in \{0, 1\}$. In NoiseMod, the modulation alphabet is defined by the variance of the transmitted complex Gaussian signal.
Under hypothesis $\mathcal{H}_0$, the transmitter is silent, and the receiver observes only the complex background thermal noise. Under hypothesis $\mathcal{H}_1$, the transmitter emits a pseudo-random circularly symmetric complex Gaussian noise signal of total average power $P$.

Assuming perfect symbol synchronization, the complex baseband received signal $y_n \in \mathbb{C}$ at the $n$-th discrete sample within a symbol period is modeled as
\begin{align}
    \mathcal{H}_0 : y_n &= w_n, \\
    \mathcal{H}_1 : y_n &= x_n + w_n,
\end{align}
where $w_n \sim \mathcal{CN}(0, \sigma_0^2)$ represents the complex Additive White Gaussian Noise (AWGN), and $x_n \sim \mathcal{CN}(0, P)$ is the transmitted complex noise signal. Because the sum of independent complex Gaussian variables remains complex Gaussian, the received signal under each hypothesis is
\begin{align}
    y_n | \mathcal{H}_0 &\sim \mathcal{CN}(0, \sigma_0^2), \\
    y_n | \mathcal{H}_1 &\sim \mathcal{CN}(0, \sigma_1^2),
\end{align}
where $\sigma_1^2 = P + \sigma_0^2$. The instantaneous, per-sample Signal-to-Noise Ratio is $\gamma_\text{snr} = {P}/{\sigma_0^2}$.

\vspace{-5pt}
\subsection{Optimal Detection Threshold in Complex AWGN}

The receiver collects $N$ independent complex samples $\mathbf{y} = [y_1, y_2, \dots, y_N]^T$ during one symbol duration. The optimal energy detector calculates the test statistic $T(\mathbf{y})$ as the sum of squared magnitudes
\begin{equation}
\small
    T(\mathbf{y}) = \sum_{n=1}^{N} |y_n|^2.
\end{equation}
To determine the optimal decision boundary, we apply the Maximum Likelihood (ML) criterion (assuming equiprobable bits), setting the log-likelihood ratio (LLR) to zero, thus
\begin{equation}
\small
    \ln \frac{f(\mathbf{y} | \mathcal{H}_1)}{f(\mathbf{y} | \mathcal{H}_0)} \gtrless_{\mathcal{H}_0}^{\mathcal{H}_1} 0.
\end{equation}

The complex Gaussian probability density function (PDF) for a sample under $\mathcal{H}_i$ is $f(y_n|\mathcal{H}_i) = \exp\!\left(-{|y_n|^2}/{\sigma_i^2}\right)/{\pi\sigma_i^2}$. Using the independence of samples, the joint PDF is the product of marginals. Substituting into the LLR gives
\begin{equation}
\small
    \sum_{n=1}^{N} \left[ -\ln(\pi\sigma_1^2) - \frac{|y_n|^2}{\sigma_1^2} + \ln(\pi\sigma_0^2) + \frac{|y_n|^2}{\sigma_0^2} \right] = 0.
    \label{eq_log_sun}
\end{equation}

After straightforward manipulation, the optimal threshold for complex NoiseMod is
\begin{equation}
\small
    \gamma_\text{th} = N \frac{\sigma_0^2 \sigma_1^2}{\sigma_1^2 - \sigma_0^2} \ln\!\left( \frac{\sigma_1^2}{\sigma_0^2} \right). \label{eq:threshold}
\end{equation}
This expression is formally identical to the real-valued case, but the test statistic now aggregates the squared magnitudes of complex samples, which alters its distribution, as shown next. In practice, $\sigma_0^2$ and $\sigma_1^2$ must be estimated from a noise-only reference interval and from calibration, respectively.

\vspace{-5pt}
\subsection{Analytical BER Derivation in Complex AWGN}

The test statistic $T$ is the sum of $N$ squared magnitudes of complex Gaussian random variables. It is well-known that $2|y_n|^2/\sigma_0^2 \sim \chi^2_2$ under $\mathcal{H}_0$, and $2|y_n|^2/\sigma_1^2 \sim \chi^2_2$ under $\mathcal{H}_1$. Consequently, $
    \frac{2T}{\sigma_0^2} \Big| \mathcal{H}_0 \sim \chi^2_{2N}$, and 
    $\frac{2T}{\sigma_1^2} \Big| \mathcal{H}_1 \sim \chi^2_{2N}$.

Let $F_{\chi^2_{2N}}(z)$ denote the Cumulative Distribution Function (CDF) of a chi-square distribution with $2N$ degrees of freedom. The probabilities of false alarm ($P_\text{fa}$) and missed detection ($P_\text{md}$) are
\begin{align}
    P_\text{fa} &= \mathbb{P}(T > \gamma_\text{th} | \mathcal{H}_0) = 1 - F_{\chi^2_{2N}}\!\left( \frac{2 \gamma_\text{th}}{\sigma_0^2} \right), \label{eq:Pfa} \\
    P_\text{md} &= \mathbb{P}(T < \gamma_\text{th} | \mathcal{H}_1) = F_{\chi^2_{2N}}\!\left( \frac{2 \gamma_\text{th}}{\sigma_1^2} \right). \label{eq:Pmd}
\end{align}

The overall theoretical BER for the complex AWGN channel is therefore
\begin{equation}
\footnotesize
\begin{split}
    P_{\text{e}}^{\text{AWGN}} = \frac{1}{2} P_\text{fa} + \frac{1}{2} P_\text{md} = \frac{1}{2} \left[ 1 - F_{\chi^2_{2N}}\!\left(\frac{2 \gamma_\text{th}}{\sigma_0^2}\right) + F_{\chi^2_{2N}}\!\left(\frac{2 \gamma_\text{th}}{\sigma_1^2}\right) \right]. \label{eq:ber}
\end{split}
\end{equation}

\vspace{-15pt}
\subsection{SNR vs. $E_b/N_0$: A Fair Comparison Metric}
\label{sec:snr_vs_ebno}

Because BPSK and NC-FSK convey one bit per transmitted sample, while NoiseMod integrates $N$ samples per bit, the per-sample SNR $\gamma_\text{snr}$ used in \eqref{eq:threshold}, \eqref{eq:ber} is \emph{not} directly comparable, bit-for-bit, across schemes. The physically meaningful, bit-energy-normalized figure of merit is
\begin{equation}
\small
    \frac{E_b}{N_0} = \begin{cases} \gamma_\text{snr}, & \text{BPSK, NC-FSK (1 sample/bit)}, \\ N\cdot\gamma_\text{snr}, & \text{NoiseMod ($N$ samples/bit)}. \end{cases}
    \label{eq:ebno}
\end{equation}

This distinction matters in practice: at a \emph{fixed per-sample SNR}, integrating $N$ samples gives NoiseMod's energy detector an averaging (non-coherent combining) gain that can make its BER curve cross \emph{below} the single-sample BPSK/NC-FSK curves for large enough $N$, but this is not a ``free'' gain: it is purchased with $N\times$ more transmitted samples (and, as shown in section III-C, a correspondingly wider receiver noise bandwidth). When results are expressed in $E_b/N_0$ via \eqref{eq:ebno}, NoiseMod's genuine energy-efficiency penalty appears. All BER-vs-SNR curves in this paper use the per-sample $\gamma_\text{snr}$, which is the correct axis for characterizing detector behavior at a given instantaneous channel condition. The link-budget and energy analysis of sections III, and V-F, which determines transmit power and range, uses $E_b/N_0$.

\vspace{-6pt}
\subsection{System Model under Rayleigh Fading}
Under frequency-flat Rayleigh fading, the complex received signal becomes
\begin{align}
    \mathcal{H}_0 : y_n &= w_n, \\
    \mathcal{H}_1 : y_n &= h x_n + w_n,
\end{align}
where $h \sim \mathcal{CN}(0, 1)$ is the complex fading coefficient \cite{b10, b11}, independent of $x_n$ and $w_n$. The instantaneous channel power gain $g = |h|^2$ follows an exponential distribution $f_G(g) = e^{-g}$ for $g \ge 0$.

Conditional on a specific channel gain $g$, the received signal under $\mathcal{H}_1$ is complex Gaussian with total variance
\begin{equation}
\small
    \sigma_1^2(g) = g P + \sigma_0^2.
\end{equation}

Without instantaneous CSI, the receiver keeps the threshold $ \gamma_\text{th}$ fixed according to the average SNR. The false alarm probability $P_\text{fa}$ remains unchanged because $\mathcal{H}_0$ depends only on the local receiver noise $\sigma_0^2$. The conditional missed detection probability, however, becomes
\begin{equation}
\small
    P_\text{md}(g) = F_{\chi^2_{2N}}\!\left( \frac{2 \gamma_\text{th}}{g P + \sigma_0^2} \right). \label{eq:Pmd_fading}
\end{equation}

The average BER over the Rayleigh fading channel is obtained by integrating the conditional missed detection probability over the fading distribution:
\begin{equation}
    P_{\text{e}}^{\text{Rayleigh}} = \frac{1}{2} P_\text{fa} + \frac{1}{2} \int_{0}^{\infty} F_{\chi^2_{2N}}\!\left( \frac{2 \gamma_\text{th}}{g P + \sigma_0^2} \right) e^{-g} \, \text{d}g. \label{eq:fading_ber}
\end{equation}
This integral is evaluated numerically. Deep fades (small $g$) make $\sigma_1^2(g) \approx \sigma_0^2$, dramatically increasing the missed detection rate and causing error floors, as shown in Section V-D.

\vspace{-5pt}
\subsection{Selection Diversity: Idealized Bound and a Practical, CSI-Free Scheme}
\label{sec:diversity}

Without CSI, the average BER exhibits a pronounced high-SNR performance degradation under Rayleigh fading. To establish a theoretical performance bound, we first analyze an idealized 2-antenna selection-diversity scheme in which branch selection uses perfectly known instantaneous channel gains, $g_{\max} = \max(g_1, g_2)$, with PDF $f(g_{\max}) = 2(1 - e^{-g})e^{-g}$. The idealized, diversity-enhanced BER lower bound is then
\begin{equation}
\footnotesize
P_\text{e}^\text{div,ideal} = \frac{1}{2} P_\text{fa} + \frac{1}{2} \int_{0}^{\infty} F_{\chi^2_{2N}}\!\left( \frac{2 \gamma_\text{th}}{g P + \sigma_0^2} \right) 2(1 - e^{-g})e^{-g} \, \text{d}g. \label{eq:fading_div}
\end{equation}

A practical CSI-free implementation selects the antenna branch with the largest integrated energy,
$T_{\rm sel}=\max(T_1,T_2)$,
and applies the same threshold $\gamma_\text{th}$. Because $T_1,T_2$ are i.i.d.\ under $\mathcal{H}_0$, the resulting false-alarm probability is
\begin{equation}
\small
    P_\text{fa}^\text{prac} = 1-\Big[F_{\chi^2_{2N}}\!\left(2\gamma_\text{th}/\sigma_0^2\right)\Big]^2.
    \label{eq:pfa_practical}
\end{equation}

The missed-detection probability does not admit a simple closed form (it involves the order statistics of two correlated-by-selection, independently faded branches) and is evaluated by Monte Carlo simulation. As shown in Fig. \ref{fig:4}, the practical scheme closely follows the idealized bound \eqref{eq:fading_div}, demonstrating that near-optimal diversity can be achieved without channel estimation. This is, to our knowledge, a new practical result for this detector.

We restrict the diversity analysis to NoiseMod because it is the scheme that exhibits the most pronounced degradation under Rayleigh fading without CSI and is therefore the primary beneficiary of the proposed diversity technique. BPSK and NC-FSK do not suffer this floor under our AWGN assumptions and can, in principle, also benefit from classical (coherent or non-coherent) diversity combining, which is a separate, well-studied problem \cite{b10} outside the present scope.

\vspace{-5pt}
\section{Capacity and Energy Consumption Analysis}

\subsection{Channel Capacity Comparison}
We model each link as a Binary Symmetric Channel (BSC) whose crossover probability is the scheme's BER at the operating SNR. For a BSC with crossover probability $p$, the capacity per channel use is $C(p) = 1 - H_b(p)$, where $H_b(p)=-p\log_2 p - (1-p)\log_2(1-p)$ is the binary entropy function. Since BPSK and NC-FSK use one complex sample per bit,
\begin{align}
    C_\text{bpsk}(\gamma_\text{snr}) = 1-H_b\!\left(P_e^\text{bpsk}\right), \\
    C_\text{fsk}(\gamma_\text{snr}) = 1-H_b\!\left(P_e^\text{fsk}\right),
\end{align}
expressed in bits/channel-use (equivalently, bits/sample). NoiseMod, however, spends $N$ complex samples to convey one bit, so its capacity \emph{per sample}, the resource shared with BPSK/NC-FSK, is
\begin{equation}
\small
    C_\text{nm}(\gamma_\text{snr},N) = [1-H_b\!\left(P_e^\text{AWGN}(\gamma_\text{snr},N)\right)]/N.
    \label{eq:capacity_nm}
\end{equation}
Because $P_e^\text{AWGN}(\gamma_\text{snr},N)\to 0$ for a fixed $N$ as $\gamma_\text{snr}$ grows, $C_\text{nm}$ saturates at $1/N$ bits/sample rather than at 1 bit/sample: for $N=50$, the ceiling is $1/50=0.02$ bits/sample. This capacity comparison is already fair on a per-channel-use basis, since $C_\text{nm}$ is explicitly normalized by $N$.

\vspace{-5pt}
\subsection{Energy Consumption Estimation}
To evaluate hardware viability, we estimate the total energy consumed per bit, $E_\text{bit}$. Let $R_\text{b}$ be the target transmission rate (bits/s). With $P_\text{TX\_ckt}$ and $P_\text{RX\_ckt}$ the baseline circuitry power of the transmitter and receiver, $P_\text{PA}$ the radiated RF power, and $\eta_\text{PA}$ the power-amplifier drain efficiency,
\begin{equation}
\small
    E_\text{bit} = \frac{P_\text{TX\_ckt}}{R_\text{b}} + \frac{P_\text{PA}}{\eta_\text{PA} R_\text{b}} + \frac{P_\text{RX\_ckt}}{R_\text{b}}.
    \label{eq:ebit_simple}
\end{equation}

Traditional coherent/non-coherent schemes such as BPSK and FSK rely on local oscillators (LO), Phase-Locked Loops (PLL), and active mixers. Published ultra-low-power BLE/IEEE~802.15.4-class transceivers report circuit power in the sub-mW to few-mW range depending on process node and design generation, e.g., a 3.7~mW transmitter with a 2.75~mW receiver in 28-nm CMOS \cite{b16}, with more recent designs reporting sub-2~mW receivers. We adopt $P_\text{TX\_ckt}\approx 1$~mW and $P_\text{RX\_ckt}\approx 1.5$~mW as representative, moderately optimistic values within this published range (combined circuit power $2.5$~mW).

NoiseMod replaces the frequency synthesizer with an amplified thermal-noise diode ($P_\text{TX\_ckt}\approx0.1$~mW) and the coherent front-end with a square-law envelope detector and integrator ($P_\text{RX\_ckt}\approx0.5$~mW) \cite{b13}, for a combined circuit power of $0.6$~mW, about $4\times$ lower than the coherent baseline. As shown next, this hardware saving must be weighed against the $E_b/N_0$ penalty quantified in section \ref{sec:snr_vs_ebno}.

\vspace{-5pt}
\subsection{ADC-Aware Energy Estimation Model}

Since NoiseMod integrates $N$ samples per bit, its ADC sampling frequency must be $f_s = N \cdot R_\text{b}$, $N$ times higher than the $f_s=R_b$ needed by single-sample-per-bit BPSK/NC-FSK. This has two consequences that are easy to conflate: the receiver noise bandwidth, and hence the absolute noise power $\sigma_0^2 = N_0\cdot f_s$, scales with $N$, while the \emph{per-sample} SNR required for a target BER decreases with $N$. The two effects combine so that the required $E_b/N_0$, and hence the required absolute receive power for a fixed $R_b$, is only weakly dependent on $N$. We partition $P_\text{RX\_ckt}$ into a static baseband power $P_\text{BB}$ and a dynamic ADC energy term
\begin{equation}
\small
    E_\text{bit} = \frac{P_\text{TX\_ckt}}{R_\text{b}} + \frac{P_\text{PA}}{\eta_\text{PA} R_\text{b}} + \frac{P_\text{BB}}{R_\text{b}} + N \cdot E_\text{ADC},
    \label{eq:ebit_adc}
\end{equation}
where $E_\text{ADC}$ is the energy per sample-conversion. This reveals a three-way trade-off: improving BER by increasing $N$ raises ADC energy linearly, while improving it by increasing $\gamma_\text{snr}$ raises radiated power exponentially with distance. For BPSK/NC-FSK, which sample at the (comparatively low) bit rate itself, we retain the lumped $P_\text{RX\_ckt}$ of \eqref{eq:ebit_simple}, since their fixed, narrow-bandwidth sampling requirement does not introduce the same $N$-dependent scaling.

\vspace{-3pt}
\section{Proposed Study and Simulation Scenarios}

\vspace{-3pt}
\subsection{Definition of Scenarios}

The BER performance (Figs.~\ref{fig:1}, \ref{fig:4}) is evaluated over a per-sample SNR range of $-5$ to $20$ dB, representative of low-power sensor networks, for sample sizes $N \in \{10, 50, 100\}$; larger $N$ improves BER at a fixed per-sample SNR but reduces the bit rate and increases ADC activity, as discussed in section III-C. The link-budget/energy analysis of section V-F uses $N=50$ as its headline case, consistent with Fig.~\ref{fig:1}, \ref{fig:1b}, and additionally reports a sensitivity sweep over $N\in\{10,20,50,100\}$ (Table~\ref{tab:nsens}) showing that the energy crossover distance changes only mildly with $N$, for the reason explained in section III-C. Performance is analyzed under both AWGN (static line-of-sight) and Rayleigh fading channels, the latter modeling severe multipath conditions with complex channel coefficient $h \sim \mathcal{CN}(0, 1)$.

\vspace{-5pt}
 \subsection{Comparison Baselines}

NoiseMod is benchmarked against two established communication standards. The first is Coherent BPSK, the gold standard for power efficiency, with BER $P_{\text{e}}^\text{bpsk} = Q(\sqrt{2\gamma_\text{snr}})$; this requires carrier-phase recovery via a PLL. A differentially encoded variant (DBPSK) removes the need for absolute phase recovery, at a small (${\approx}3$~dB) performance cost, and is a natural intermediate point between full coherent BPSK and NoiseMod in the complexity-performance space \cite{b1}.

The second baseline is Non-Coherent FSK (NC-FSK), used in low-power IoT protocols such as Bluetooth Low Energy (BLE), with BER $P_{\text{e}}^\text{fsk} = \exp(-{\gamma_\text{snr}}/{2})/{2}$. NC-FSK avoids phase synchronization but still requires a stable frequency oscillator.

\vspace{-6pt}
\section{Numerical Results}

\subsection{Performance in AWGN Channel: Per-Sample SNR}
\begin{figure}[b!]
\vspace{-20pt}
\centering
\includegraphics[width=0.75\linewidth]{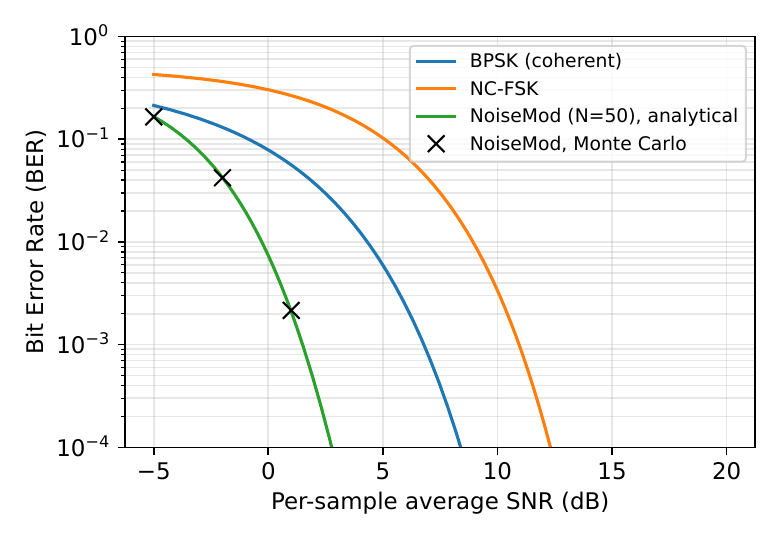}
\vspace{-10pt}
\caption{BER vs. per-sample SNR in AWGN, NoiseMod ($N=50$) vs. Coherent BPSK and Non-Coherent FSK. Markers: Monte Carlo ($4\times10^5$ trials/point).}
\label{fig:1}
\vspace{-12pt}
\end{figure}

Fig.~\ref{fig:1} compares the BER of NoiseMod ($N=50$) against BPSK and NC-FSK versus the per-sample SNR $\gamma_\text{snr}$, with Monte Carlo points confirming the analytical curves. Perhaps counter-intuitively, NoiseMod's curve crosses \emph{below} both references at moderate-to-high SNR: at a target BER of $10^{-3}$, NoiseMod needs a per-sample SNR of only $1.5$~dB, versus $6.8$~dB for BPSK and $10.9$~dB for NC-FSK. As explained in section \ref{sec:snr_vs_ebno}, this is the result of the receiver integrating $N=50$ independent samples for every bit, an averaging gain that is real, but is bought with $50\times$ more channel uses (and receiver bandwidth) per bit than BPSK/NC-FSK. It is \emph{not} evidence that NoiseMod is more energy-efficient; that comparison requires the $E_b/N_0$ axis of Fig.~\ref{fig:1b}.

\subsection{Performance in AWGN Channel: $E_b/N_0$}
\begin{figure}[htbp]
\vspace{-8pt}
\centering
\includegraphics[width=0.75\linewidth]{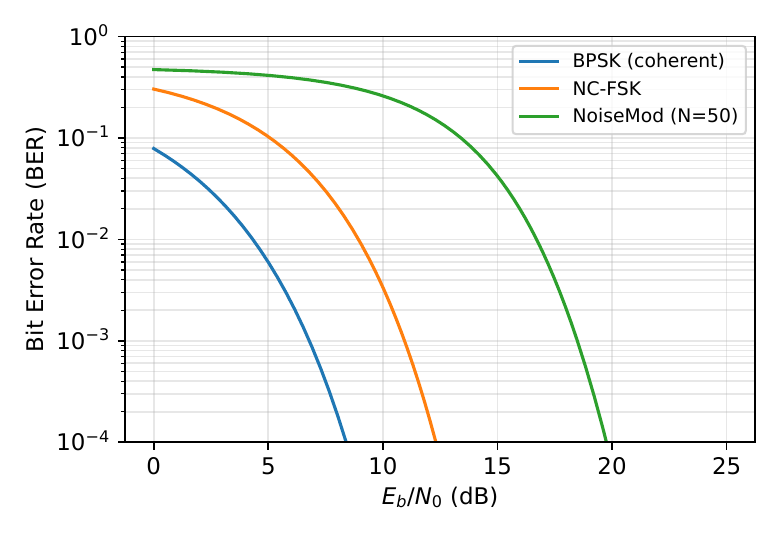}
\vspace{-10pt}
\caption{BER vs. $E_b/N_0$ in AWGN, the fair, bit-energy-normalized comparison.}
\label{fig:1b}
\vspace{-14pt}
\end{figure}

For a fair Comparison, Fig.~\ref{fig:1b} re-expresses the same three curves in terms of $E_b/N_0$ using \eqref{eq:ebno}. Here, the ranking reverses: BPSK requires $6.8$~dB, NC-FSK requires $10.9$~dB, and NoiseMod ($N=50$) requires $18.5$~dB of $E_b/N_0$ for a $10^{-3}$ BER, a penalty of approximately $7.5$~dB relative to NC-FSK and $11.7$~dB relative to BPSK. This is the correct metric to be used whenever comparing transmit-energy requirements or deriving a link budget. The required $E_b/N_0$ is only weakly sensitive to $N$: it varies from $18.2$~dB ($N=10$) to $19.3$~dB ($N=100$), i.e., by about 1~dB across a full decade of $N$ (see also Table~\ref{tab:nsens}).

\vspace{-5pt}
\subsection{Impact of Sample Size Constraints}
\begin{figure}[b!]
\vspace{-20pt}
\centering
\includegraphics[width=0.75\linewidth]{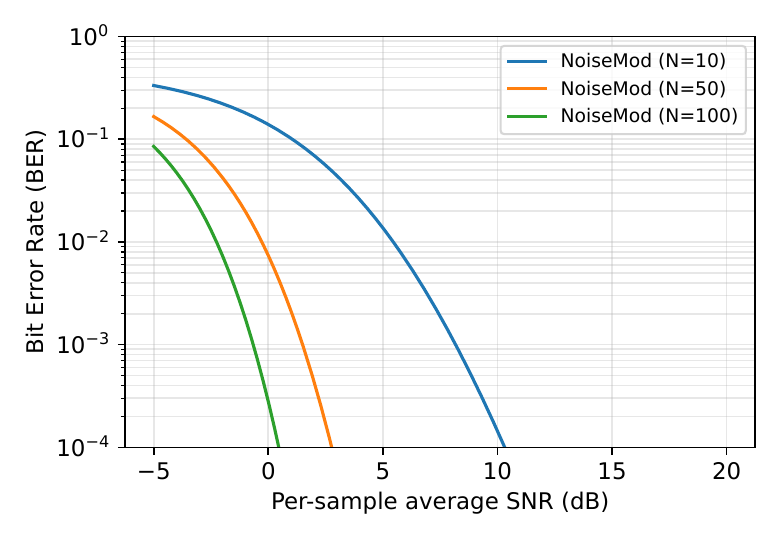}
\vspace{-8pt}
\caption{NoiseMod BER vs. per-sample SNR for $N \in \{10, 50, 100\}$.}
\label{fig:2}
\vspace{-12pt}
\end{figure}

Fig.~\ref{fig:2} shows NoiseMod's per-sample-SNR BER for $N\in\{10,50,100\}$. Increasing $N$ monotonically lowers the per-sample SNR needed for a given BER (better variance estimation), but, as quantified in sections III-C and V-F, this does not translate into a proportionally lower energy cost, since both the required $E_b/N_0$ and the ADC energy term $N\cdot E_\text{ADC}$ grow (mildly, and linearly, respectively) with $N$. Larger $N$ also linearly decreases the symbol (bit) rate for a fixed sample rate, directly decreasing capacity.

\vspace{-5pt}
\subsection{Vulnerability to Multipath Fading and Diversity Mitigation}

\begin{figure}[htbp]
\vspace{-5pt}
\centering
\includegraphics[width=0.75\linewidth]{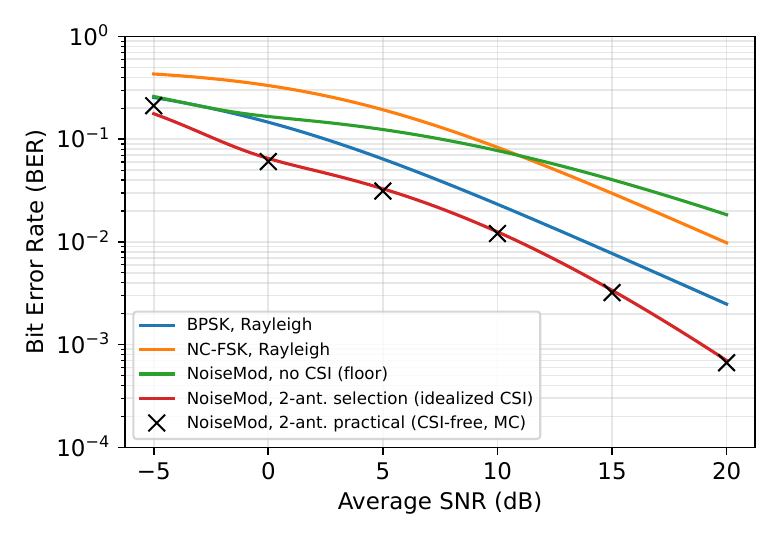}
\vspace{-10pt}
\caption{NoiseMod under Rayleigh fading: no-CSI floor, idealized 2-antenna selection diversity, and a practical CSI-free max-energy selection rule (Monte Carlo with $3\times10^5$ trials/point), with BPSK/NC-FSK Rayleigh references.}
\label{fig:4}
\vspace{-10pt}
\end{figure}

Fig.~\ref{fig:4} shows NoiseMod's dramatic degradation under flat Rayleigh fading (Eq.~\eqref{eq:fading_ber}), alongside Rayleigh-fading BPSK and NC-FSK references for context. Unlike coherent/non-coherent systems that combat fading through equalization or diversity, a static-threshold, non-CSI energy detector cannot distinguish a deep fade from a transmitted logical 0, producing a BER floor around $1$--$2\%$ even at $20$~dB average SNR, well above any practically useful operating point. The idealized 2-antenna selection-diversity bound (Eq.~\eqref{eq:fading_div}) restores a waterfall-like behavior. Critically, the fully practical, CSI-free max-energy selection rule of section \ref{sec:diversity} (Monte Carlo, markers in Fig.~\ref{fig:4}) tracks this idealized bound closely across the entire SNR range, indicating that the diversity gain is achievable without any channel-estimation overhead, an encouraging result for ULP deployments where CSI acquisition would itself consume a non-negligible energy budget.

\vspace{-5pt}
\subsection{Capacity}
\begin{figure}[b!]
\vspace{-20pt}
\centering
\includegraphics[width=0.75\linewidth]{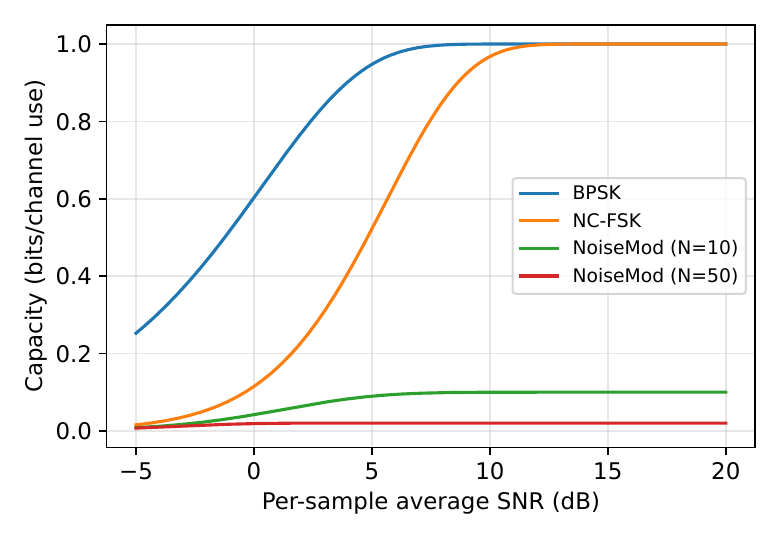}
\vspace{-10pt}
\caption{BSC-based capacity: BPSK, NC-FSK, and NoiseMod ($N=10,50$).}
\label{fig:5}
\vspace{-10pt}
\end{figure}

Fig.~\ref{fig:5} shows the BSC-based capacity comparison. BPSK and NC-FSK both approach $1$~bit/sample at high SNR (NC-FSK more slowly, reflecting its higher BER at a given SNR). NoiseMod's per-sample capacity saturates at $1/N$ bits/sample, $0.02$ for $N=50$, confirming that its practical spectral efficiency is fundamentally rate-limited by the required integration window, independent of how favorable the channel becomes.

\vspace{-5pt}
\subsection{Energy Consumption and Distance Trade-offs}

\begin{table}[t]
\centering
\caption{Link-budget parameters}
\label{tab:link}
\vspace{-5pt}
\resizebox{\columnwidth}{!}{%
\begin{tabular}{@{}ll@{}}
\toprule
Parameter & Value \\
\midrule
Target bit rate, $R_b$ & $100$ kbit/s \\
Noise reference temperature, $T_0$ & $290$ K \\
Receiver noise figure, NF & $6$ dB \\
PA drain efficiency, $\eta_\text{PA}$ & $0.30$ \\
Antenna gains, $G_t, G_r$ & $0$ dBi (isotropic) \\
Path loss model & Free-space, $\text{PL}(d)=\left(4\pi d f/c\right)^2$ \\
ADC energy, $E_\text{ADC}$ & $1$ pJ/conversion \cite{b17} \\
NoiseMod static baseband power, $P_\text{BB}$ & $0.3$ mW \\
$P_\text{TX\_ckt}$, $P_\text{RX\_ckt}$ (NoiseMod) & $0.1$ mW, $0.5$ mW \\
$P_\text{TX\_ckt}$, $P_\text{RX\_ckt}$ (BPSK/NC-FSK) & $1.0$ mW, $1.5$ mW \\
Required $E_b/N_0$ @ BER $10^{-3}$ (BPSK) & $6.79$ dB \\
Required $E_b/N_0$ @ BER $10^{-3}$ (NC-FSK) & $10.94$ dB \\
Required $E_b/N_0$ @ BER $10^{-3}$ (NoiseMod, $N=50$) & $18.49$ dB \\
\bottomrule
\end{tabular}
}
\vspace{-15pt}
\end{table}

Table~\ref{tab:link} summarizes the parameters adopted in the energy model, while Table~\ref{tab:nsens} reports the crossover distance between NoiseMod and coherent BPSK for all considered ISM bands, sample sizes, and AWGN channel (no diversity). At short distances, circuit power dominates the total energy consumption. Owing to its oscillator-free architecture, NoiseMod achieves a substantially lower energy floor than BPSK despite its higher ADC sampling rate. As distance increases, however, the transmit power required to compensate for path loss becomes the dominant energy component. Since NoiseMod requires approximately $11.7$~dB more $E_b/N_0$ than coherent BPSK to achieve a BER of $10^{-3}$, its transmit energy increases more rapidly with distance, leading to a crossover point beyond which BPSK becomes more energy-efficient.

The results show that the crossover distance strongly depends on the carrier frequency because of the increased free-space path loss at higher frequencies. For $N=20$, the crossover distance decreases from approximately $829$~m at $2.4$~GHz to $348$~m at $5.725$~GHz and only $83$~m at $24$~GHz. In contrast, the influence of the integration window is relatively small: varying $N$ from $10$ to $100$ changes the crossover distance by less than about $17\%$ for all frequencies, reflecting the weak dependence of the required $E_b/N_0$ on $N$ as discussed in Section~III-C. These results indicate that carrier frequency has a much stronger impact on the practical operating range than the integration length, making NoiseMod particularly attractive for short-range, low-frequency ultra-low-power IoT applications, while coherent BPSK remains preferable for longer links or higher-frequency bands.

\begin{table}[!t]
\centering
\caption{Sensitivity of the crossover distance to the integration window $N$ for different ISM bands.}
\vspace{-5pt}
\label{tab:nsens}
\scriptsize
\begin{tabular}{ccccc}
\toprule
\textbf{$N$} &
\textbf{Required $E_b/N_0$} &
\textbf{2.4 GHz} &
\textbf{5.725 GHz} &
\textbf{24 GHz} \\
&
\textbf{(dB)} &
\textbf{(m)} &
\textbf{(m)} &
\textbf{(m)} \\
\midrule
10  & 18.17 & 801.18 & 335.87 & 80.12 \\
20  & 17.90 & 829.06 & 347.55 & 82.91 \\
50  & 18.49 & 770.07 & 322.82 & 77.01 \\
100 & 19.34 & 692.52 & 290.31 & 69.25 \\
\bottomrule
\end{tabular}
\vspace{-20pt}
\end{table}

\begin{figure}[t]
\vspace{-20pt}
\centering
\includegraphics[width=0.78\linewidth]{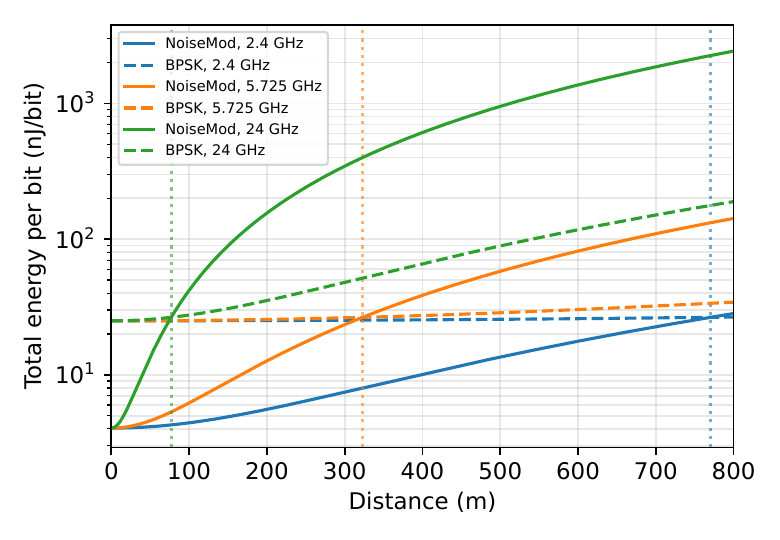}
\vspace{-10pt}
\caption{Total energy per bit vs. distance for three ISM bands (solid: NoiseMod, dashed: BPSK; $N=50$ headline case; dotted vertical lines: crossover distance).}
\label{fig:6}
\vspace{-14pt}
\end{figure}

Fig. \ref{fig:6} illustrates the total energy per bit as a function of communication distance for the three considered ISM bands and AWGN channel. At short distances, the total energy is dominated by the static circuit consumption, allowing NoiseMod to achieve a lower energy floor than coherent BPSK owing to its oscillator-free architecture and simplified receiver. As the communication distance increases, however, the transmit-power component grows rapidly because of free-space path loss. Since NoiseMod requires approximately $11.7$~dB more $E_b/N_0$ than coherent BPSK to achieve a BER of $10^{-3}$, its transmit-energy requirement increases more rapidly, producing a crossover distance beyond which BPSK becomes the more energy-efficient solution. The figure also shows that this crossover shifts to much shorter distances as the carrier frequency increases, reflecting the stronger propagation loss at higher frequencies. Consequently, NoiseMod is best suited to short-range, ultra-low-power IoT applications, whereas coherent BPSK becomes preferable for longer communication ranges, particularly in the higher-frequency ISM bands.

\vspace{-5pt}
\section{Discussion: Limitations and Trade-offs}
\label{sec:discussion}

\subsubsection{Per-Sample SNR vs. $E_b/N_0$}
As emphasized throughout section \ref{sec:snr_vs_ebno}--V, any comparison of NoiseMod against single-sample-per-bit schemes must fix which quantity is being held constant. Per-sample-SNR comparisons (as in Fig.~\ref{fig:1}) are informative about the detector's operating point but must not be read as an energy-efficiency comparison, for which $E_b/N_0$ (Fig.~\ref{fig:1b}) is required.

\subsubsection{Threshold Estimation Sensitivity}
The optimal threshold \eqref{eq:threshold} is a function of $\sigma_0^2$ and $\sigma_1^2$, both of which must be estimated from finite reference intervals in practice, whereas BPSK and (non-coherent) FSK detection does not require an analogous, channel-noise-dependent decision threshold. An error in the estimated noise floor shifts $\gamma_\text{th}$ and asymmetrically inflates $P_\text{fa}$ or $P_\text{md}$; robust, adaptive noise-floor tracking is therefore an important, and currently unaddressed, implementation requirement for NoiseMod, and a natural direction for future work.

\subsubsection{Bandwidth and Interference Vulnerability}
Because NoiseMod's front end must sample at $f_s=N R_b$ (section III-C), its noise (and interference) collection bandwidth is $N\times$ wider than that of BPSK or NC-FSK operating at the same bit rate. This makes NoiseMod inherently more exposed to in-band interference and makes coexistence in congested ISM bands more challenging than for the narrowband coherent/non-coherent alternatives, independent of the thermal-noise trade-offs already discussed.

\subsubsection{Throughput vs. Detection Reliability}
Estimating variance makes NoiseMod inherently a low-data-rate scheme: high throughput requires small $N$, which degrades both detection reliability (Fig.~\ref{fig:2}) and capacity (Fig.~\ref{fig:5}).

\subsubsection{Fading Vulnerability and Diversity}

Without instantaneous CSI, Rayleigh fading causes a pronounced degradation in BER performance at high SNR. We show that a practical, CSI-free 2-antenna selection rule closes most of this gap relative to an idealized bound, but extending this analysis to more than two branches, to correlated fading, and to a full energy accounting of the antenna front-end itself remains future work.

\vspace{-8pt}
\section{Conclusion}

This paper presented a practical assessment of NoiseMod for ultra-low-power IoT communications through a unified analytical and simulation framework. Although NoiseMod benefits from low-complexity hardware and can outperform conventional schemes on a per-sample SNR basis, a fair comparison in terms of $E_b/N_0$ reveals a significant energy penalty relative to coherent BPSK and non-coherent FSK. Moreover, the required oversampling fundamentally limits its spectral efficiency and increases its vulnerability to Rayleigh fading. We further showed that a practical CSI-free antenna-selection strategy nearly matches ideal selection diversity, effectively mitigating fading without channel estimation. Finally, our reproducible energy model demonstrates that NoiseMod is advantageous only for short-range links, where its hardware simplicity outweighs its transmit-energy penalty, with the crossover distance strongly dependent on carrier frequency. These findings provide practical design guidelines for NoiseMod-based ultra-low-power IoT systems and motivate future research on adaptive threshold estimation, interference resilience, and advanced diversity techniques.


\begin{thebibliography}{00}
\bibitem{b1} A. Al-Fuqaha et al., ``Internet of Things: A Survey on Enabling Technologies, Protocols, and Applications,'' \textit{IEEE Communications Surveys \& Tutorials}, vol. 17, no. 4, pp. 2347-2376, 2015.
\bibitem{b2} V. Liu et al., ``Ambient backscatter: wireless communication out of thin air,'' \textit{ACM SIGCOMM}, 2013.
\bibitem{b5} E. Basar, ``Communication by Means of Thermal Noise: Toward Networks With Extremely Low Power Consumption,'' \textit{IEEE Trans. Commun.}, vol. 70, no. 2, pp. 1435-1444, 2022.
\bibitem{b6} Alshawaqfeh, Mustafa K. et al. ``Thermal noise modulation: Optimal detection and performance analysis,'' \textit{IEEE Communications Letters}, vol. 28, no. 12, pp. 2930-2934, 2024.
\bibitem{b7} A. A. d. Anjos and H. S. Silva, ``On–Off Digital Noise Modulation,'' \textit{IEEE Wireless Communications Letters}, vol. 14, no. 11, pp. 3595-3599, Nov. 2025
\bibitem{b8} M. A. ElMossallamy et al., ``Noncoherent Frequency-Shift Keying for Ambient Backscatter Over OFDM Signals,'' \textit{IEEE Open Journal of the Communications Society}, vol. 5, pp. 5219-5231, 2024.
\bibitem{b9} H. T. P. Da Silva et al., ``A Survey on Noise-Based Communication,'' \textit{IEEE Access}, vol. 14, pp. 14722-14734, 2026.
\bibitem{b10} Liu, Yuan et al., ``Optimal Thresholds for Differential Energy Detection of Ambient Backscatter Communication,'' \textit{International Conference on Internet of Things as a Service}. Cham: Springer International Publishing, 2020.
\bibitem{b11} M. K. Alshawaqfeh et al., ``Fixed-Threshold Detection Strategy for Thermal Noise Modulation under Rayleigh Fading Channels,'' \textit{2025 IEEE 36th International Symposium on Personal, Indoor and Mobile Radio Communications (PIMRC)}, Istanbul, Turkiye, 2025.
\bibitem{b13} R. A. Tasci et al., ``Flip-KLJN: Random Resistance Flipping for Noise-Driven Secure Communication,'' \textit{IEEE Transactions on Communications}, vol. 73, no. 11, pp. 12625-12636, Nov. 2025.
\bibitem{b14} Jbari, Mohamed El et al. ``Optimal Bit Detection in Thermal Noise Communication Systems Under Rician Fading,'' arXiv preprint arXiv:2511.21649, 2025.
\bibitem{b15} H. Zayyani et al., ``Spread Spectrum Noise Modulation: Analysis and Detection,'' \textit{IEEE Communications Letters}, vol. 30, pp. 1106-1110, 2026.
\bibitem{b16} F.-W. Kuo et al., ``A Bluetooth Low-Energy Transceiver With 3.7-mW All-Digital Transmitter, 2.75-mW High-IF Discrete-Time Receiver, and TX/RX Switchable On-Chip Matching Network,'' \textit{IEEE J. Solid-State Circuits}, vol. 52, no. 4, pp. 1144-1162, Apr. 2017.
\bibitem{b17} X. Tang et al., ``Low-Power SAR ADC Design: Overview and Survey of State-of-the-Art Techniques,'' \textit{IEEE Trans. Circuits Syst. I: Regular Papers}, vol. 69, no. 6, pp. 2249-2262, Jun. 2022.
\end{thebibliography}
\end{document}